\documentclass[3p,times,twocolumn]{elsarticle}

\usepackage{ecrc}
\usepackage{hepnames}

\volume{00}

\firstpage{1}

\journalname{Nuclear Physics B Proceedings Supplement}

\runauth{P. Azzurri}

\jid{nuphbp}

\jnltitlelogo{Nuclear Physics B Proceedings Supplement}

\usepackage{amssymb}

\usepackage[figuresright]{rotating}

\newcommand{\ttbar}{\ensuremath{\rm{t}\overline{\rm{t}}}\xspace} 
\newcommand{\pt}{\ensuremath{p_{\mathrm{T}}}\xspace}
\newcommand{\HT}{\ensuremath{H_{\mathrm{T}}}\xspace}

\begin{document}

\begin{frontmatter}



\dochead{}

\title{Measurements of the hadronic activity and the electroweak production in events with a Z boson and two jets in proton-proton collisions with the CMS experiment }


\author{Paolo Azzurri, for the CMS Collaboration}

\address{INFN Pisa, Largo Pontecorvo 3, 56127 Pisa, Italy}

\begin{abstract}
The observation of the electroweak production of a Z boson with two
jets in pp collisions at $\sqrt{s} = 8$~TeV with the CMS experiment at the CERN LHC
is presented, based on a data sample with an integrated luminosity of 19.7~fb$^{-1}$.
The cross section measurement, combining the muon and electron channels, is in agreement with the 
theoretical expectations.
Radiation patterns of selected Z plus two jets events, and the hadronic activity in 
the rapidity interval between the jets
are also measured. These results are of substantial importance 
in the more general study of vector boson fusion processes, of relevance
for Higgs boson searches and for measurements of electroweak
gauge couplings and vector boson scattering.
\end{abstract}




\end{frontmatter}


\section{Introduction\label{sec:intro}}

In proton collisions at the LHC Vector Boson Fusion (VBF) 
happens when a valence quark of each one of the colliding protons radiates a $\PW$ or 
Z boson that subsequently interact or ``fuse''.
For both valence quark radiating the weak bosons a 
$t$-channel four-momentum with $Q^2 \sim m^2_Z,m^2_W$ is exchanged. 
In this way the two valence quarks are typically scattered
away from the beam line and inside the detector acceptance, where they can be revealed as hadronic jets. 
The distinctive signature of VBF is therefore the presence of these two energetic hadronic 
jets (tagging jets), 
roughly in the forward and backward direction with respect to the proton beam line.
 
The VBF production has a great prominence at the LHC for its importance 
for the measurements of the Higgs sector couplings~\cite{Zeppenfeld:2000td,Duhrssen:2004cv}.
The study of the VBF production of Z or W bosons is therefore an important benchmark to cross-check 
and validate Higgs VBF measurements~\cite{Khoze:2002fa}, but serves further
as a probe of triple-gauge-boson couplings~\cite{Baur:1993fv},
for searches for physics beyond the standard model~\cite{Cho:2006sx,Dutta:2012xe},
and as a precursor to the measurement of elastic vector boson pair scattering.

On the other hand the VBF production of Z or W bosons has some intriguing differences 
with respect to the Higgs VBF productions. 
When focusing on VBF Z/W production, the observed final state is composed 
of a pair of fermions (ff), either quarks or leptons, from the Z/W decay, 
associated with a pair of quarks (qq) 
from the VBF production mechanism; but in this context there is a large number of
non-VBF diagrams that lead to identical ffqq final states  
that can't be neglected~\cite{Oleari:2003tc}.

\begin{figure*}[htb] {
\centering
\includegraphics[width=0.195\textwidth]{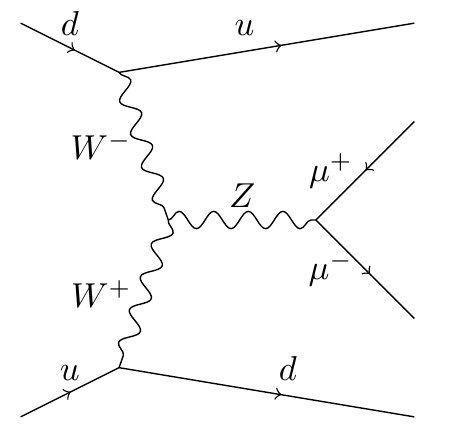}
\includegraphics[width=0.195\textwidth]{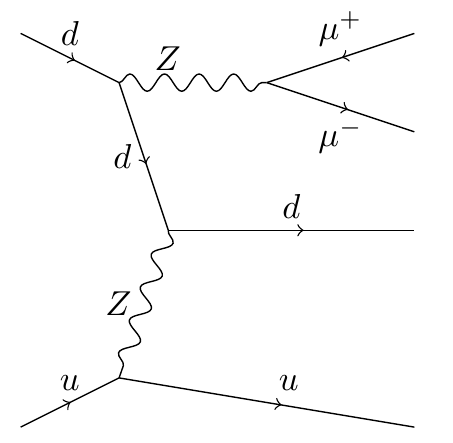}
\includegraphics[width=0.195\textwidth]{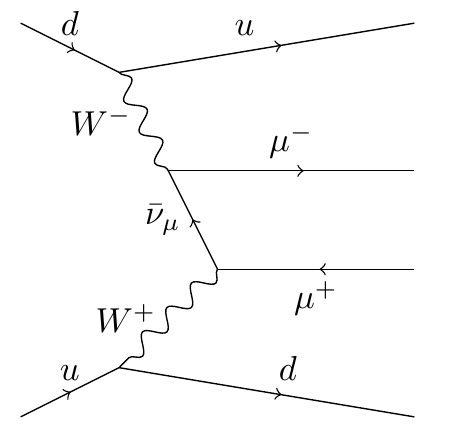}
\includegraphics[width=0.195\textwidth]{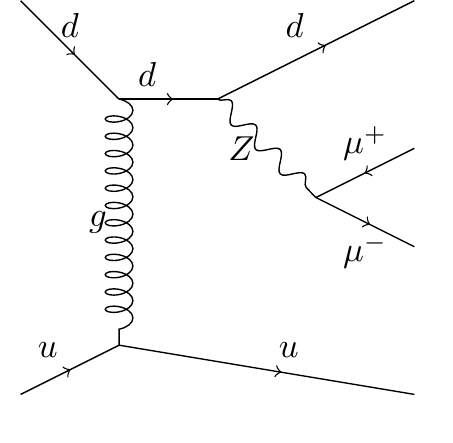}
\includegraphics[width=0.195\textwidth]{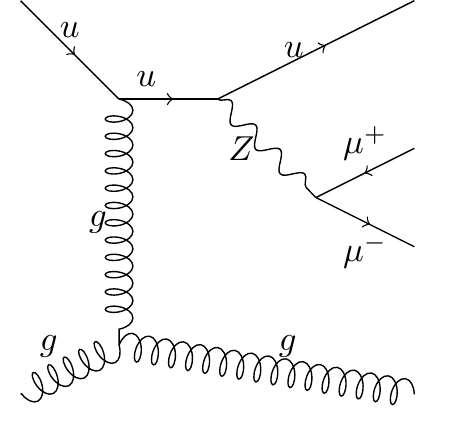}
\caption{
Representative Feynman diagrams for dimuon plus dijet productions in proton-proton 
collisions.  
Electroweak contributions:
(far left) vector boson fusion,
(center-left) Z-strahlung,
(center) multiperipheral production.
Mixed electroweak and strong contributions:
(center-right) identical to electroweak, 
(far right) with different initial and final state.
\label{fig:diagrams}}
}
\end{figure*}

Considering only the classes of
diagrams with purely electroweak (EW) interactions, (like the
VBF one),  shown in  Figure~\ref{fig:diagrams}, and no QCD interactions, the
additional diagrams have strong negative interferences with the VBF
productions. 
These large negative interference effects are in fact related to well-known non-abelian gauge cancellations 
that preserve the scattering unitarity and the electroweak model 
theory renormalizability~\cite{'tHooft:1972fi}. 
This situation makes the VBF Z/W channel more complicated but also more interesting. 

Another main scope of selecting ``VBF-like'' Z plus two jets events, is to 
study the event hadronization properties 
connected with the peculiar VBF production color structure.
In VBF processes and more in general also for the contributing electroweak processes with identical
final states, there is no $t$-channel color exchange. 
This leads to the expectation of a ``rapidity gap'' of suppressed hadronic activity between the two 
tagging jets that is a very peculiar feature, in particular in 
the case of a large rapidity separation between the two 
tagging jets~\cite{Chehime:1992ub,Bjorken:1992er,Rainwater:1996ud}.
Measurements of the additional hadronic activity in the rapidity gap provides a precious validation
of the Monte Carlo models simulations and benchmark results 
for the use of rapidity gap observables, like jet vetoes, 
in independent VBF event productions (e.g. for Higgs selections).

At the LHC, the EW Z plus two jets process was first measured by the CMS 
experiment using pp collisions at
$\sqrt{s}=7$~TeV~\cite{Chatrchyan:2013jya},
and more recently by both the ATLAS and CMS experiments
with $\sqrt{s}=8$~TeV data~\cite{Aad:2014dta,Khachatryan:2014dea}.
The results presented here will focus on the most recent CMS results 
using pp collisions collected at $\sqrt{s}=8$~TeV and
corresponding to an integrated luminosity of 19.7~fb$^{-1}$~\cite{Khachatryan:2014dea}.
Different methods have been used 
to confirm the presence of the signal: 
two multivariate analyses methods (A) and (B) as developed for the 7~TeV 
analysis~\cite{Chatrchyan:2013jya}, 
and new method (C) with a data-driven  
model of the main Drell-Yan (DY) background.

\section{Event reconstruction and simulation}
A detailed description of the CMS detector can be found in Ref.~\cite{Chatrchyan:2008aa}.
The central feature of the CMS apparatus is a superconducting solenoid of 6~m internal diameter
providing a field of 3.8~T. Within the field volume are a silicon pixel and strip tracker, a crystal
electromagnetic calorimeter (ECAL), and a brass/scintillator hadron calorimeter (HCAL)
providing coverage for pseudorapidities $|\eta | < 3$.
The forward calorimeter modules extend the coverage of hadronic jets up to $|\eta | < 5$.

Electrons are reconstructed from clusters of energy
depositions in the ECAL that match tracks extrapolated from the
silicon tracker~\cite{CMS-PAS-EGM-10-004}.
Muons are reconstructed by fitting trajectories based on hits in the silicon
tracker and in the outer muon system~\cite{muon}. 

Jets are clustered using the
anti-$k_{\rm T}$ algorithm~\cite{Cacciari:2008gp} with
a distance parameter of 0.5. 
Two different types of jets are used in the analysis: jet-plus-track (JPT)
and particle-flow (PF) jets. 
The JPT jets are reconstructed calorimeter jets whose energy response and resolution are improved
by incorporating tracking information according to the JPT algorithm~\cite{CMS-PAS-JME-09-002}.
The CMS particle flow algorithm~\cite{CMS-PAS-PFT-09-001,CMS-PAS-PFT-10-002} combines the information
from all relevant CMS sub-detectors to identify and reconstruct particle candidates in the event, and 
PF jets are reconstructed clustering particles identified by the particle flow algorithm.

The signal is defined as the pure EW production of $\ell\ell$jj final states in 
the kinematic region defined by dilepton mass $M_{\ell\ell} >50$~GeV,
parton transverse momentum $p_{\rm T j}  > 25$~GeV,
parton pseudorapidity $\vert \eta_{\rm j}\vert< 5$,
diparton mass $M_{\rm jj} >120$~GeV.

Signal events are simulated at leading order (LO) using
the MADGRAPH Monte Carlo (MC) generator~\cite{Alwall:2011uj,Alwall:2014hca},
interfaced to PYTHIA (v6.4.26)~\cite{Sjostrand:2006za} for parton showering (PS) and hadronisation.
The underlying event is modeled with the so-called $Z2^{*}$ tune~\cite{Chatrchyan:2011id}.
The predicted signal cross section is 
$\sigma_{\rm LO}({\rm EW}~\ell\ell{\rm jj})=208~\pm 16$~fb, for a single lepton 
flavor.

Background DY $\ell\ell$ events are also generated 
with MADGRAPH using a LO matrix element (ME) calculation that includes up to four partons
generated from quantum chromodynamics (QCD) interactions, and interfaced to PYTHIA for PS.
The ME-PS matching is
performed following the ktMLM prescription~\cite{Mangano:2006rw,Alwall:2007fs}.

Possible LO interference effects between the EW signal and DY processes
have been evaluated making use of MADGRAPH, comparing the differential 
distributions of samples with (i) pure signal, (ii) pure DY plus two partons,
and (iii) both signal and DY together.

Other residual backgrounds from ditop (\ttbar) and diboson (VV) productions 
are generated with MADGRAPH, while single top productions 
are generated with POWHEG~\cite{Alioli:2010xd}.

The CMS detector simulation, based on GEANT4~\cite{Allison:2006ve,Agostinelli:2002hh},
is applied to all the generated signal and background samples.
The presence of multiple pp interactions in the same beam crossing  (pileup) 
is incorporated by simulating additional interactions 
(both in-time and out-of-time with the collision) with a multiplicity
that matches the one observed in data.
The average number of pileup events in the 8~TeV data is estimated as $\approx$21
interactions per bunch crossing.

\section{Event selection and Drell-Yan background model}
Opposite sign lepton pairs are selected with validated CMS algorithms 
for electrons ~\cite{CMS-PAS-EGM-10-004} and muons~\cite{muon}.
A relative lepton isolation is defined as $I= \sum{{\pt}_{\text{i}}} / {\pt}_{\ell}$, 
where the sum includes all reconstructed PF objects 
inside a cone of $\Delta R = \sqrt{(\Delta\eta)^2+(\Delta\phi)^2} < 0.4$ around 
the lepton.
Each lepton is required to have a transverse momentum in excess of 20~GeV,
and a relative isolation $I$ smaller than 0.10 and 0.12 for electrons and muons
respectively. The invariant mass $M_{\ell\ell}$ of selected same flavor leptons is finally required 
to satisfy $|M_{\rm Z}-M_{\ell\ell}|<15$~GeV, where $M_{\rm Z}$ is the nominal 
Z-boson mass.

Analyses (A) and (B) make respectively use of PF and JPT reconstructed jets in the selected events, and both 
rely on MC simulations to predict the main DY plus jets background.
Analysis (C) makes use of PF jets, as analysis (A), but uses a model of DY plus jets derived from 
 photon plus jets data events~\cite{Khachatryan:2014dea}, where the requirement 
$\pt\mathrm{(Z/\gamma)}>50$~GeV is applied to ensure a good photon purity.
It is further verified that the data-driven method works correctly with simulated events.

For the rapidity gap and signal measurements events are required to have
two PF or JPT jets within $ | \eta |\leq 4.7$,
with $\pt>50, 30$~GeV and with a dijet invariant mass   
$M_\mathrm{jj}>200$~GeV for the \pt-leading and subleading jets.

\section{Event jet radiation patterns}
The selected Z plus jets event ``radiation patterns'' are studied, and for this, according to the prescriptions in 
Ref.~\cite{Binoth:2010ra}, only PF jets with $\pt>40$~GeV are considered. 
The investigated observables  are :
(i) the number of jets, $N_\mathrm{j}$,
(ii) the total scalar sum of the transverse momenta of jets
reconstructed within $\vert\eta \vert<4.7$,  $\HT$,
(iii) $\Delta\eta_\mathrm{jj}$ between the two jets 
    which span the largest pseudorapidity gap in the event, and
(iv) the cosine of the azimuthal angle difference, $\cos \Delta\phi_\mathrm{jj}$,
     for the two jets with criterion (iii).

\begin{figure}[htb] {
\centering
\includegraphics[width=0.234\textwidth]{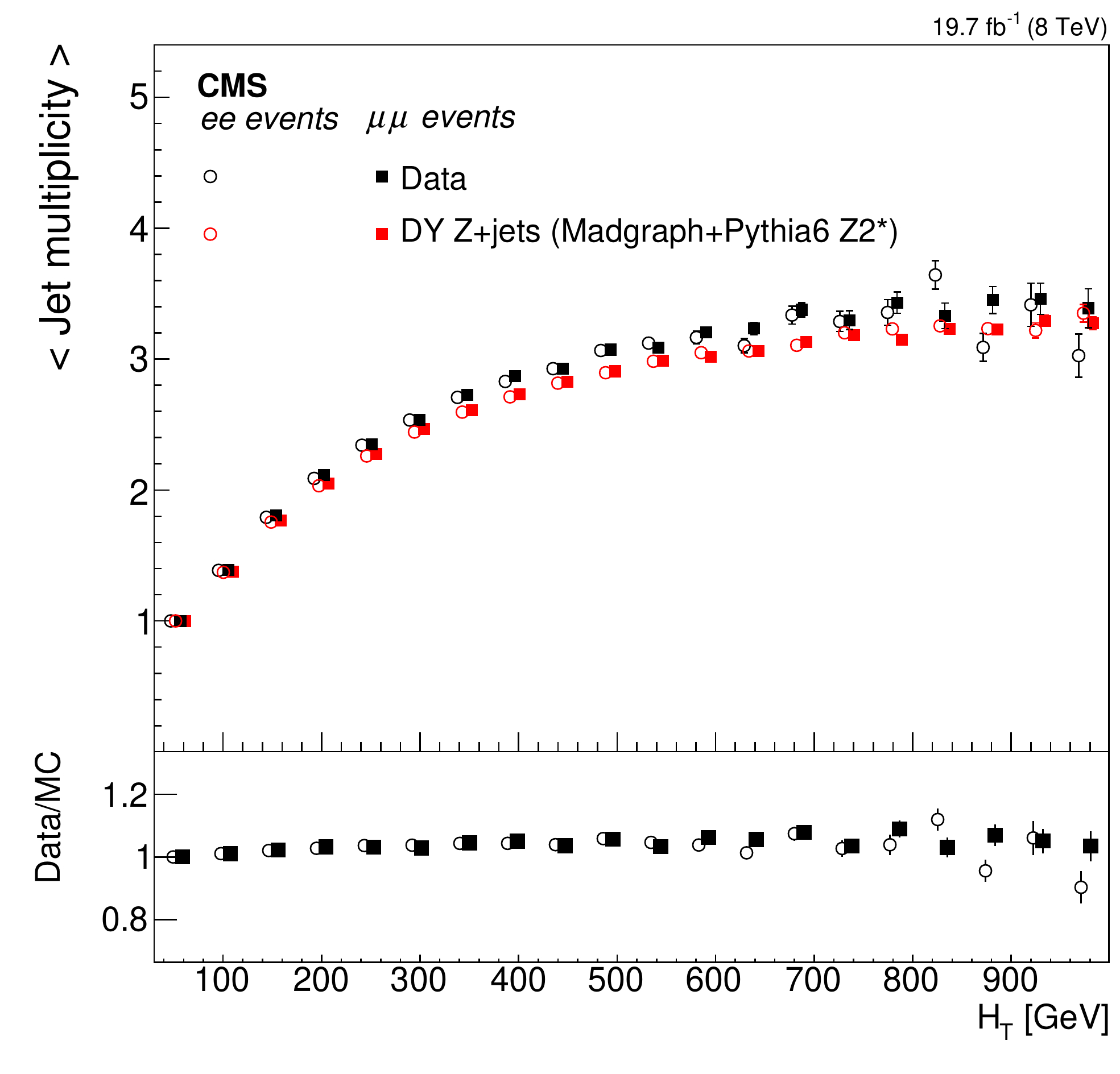}
\includegraphics[width=0.234\textwidth]{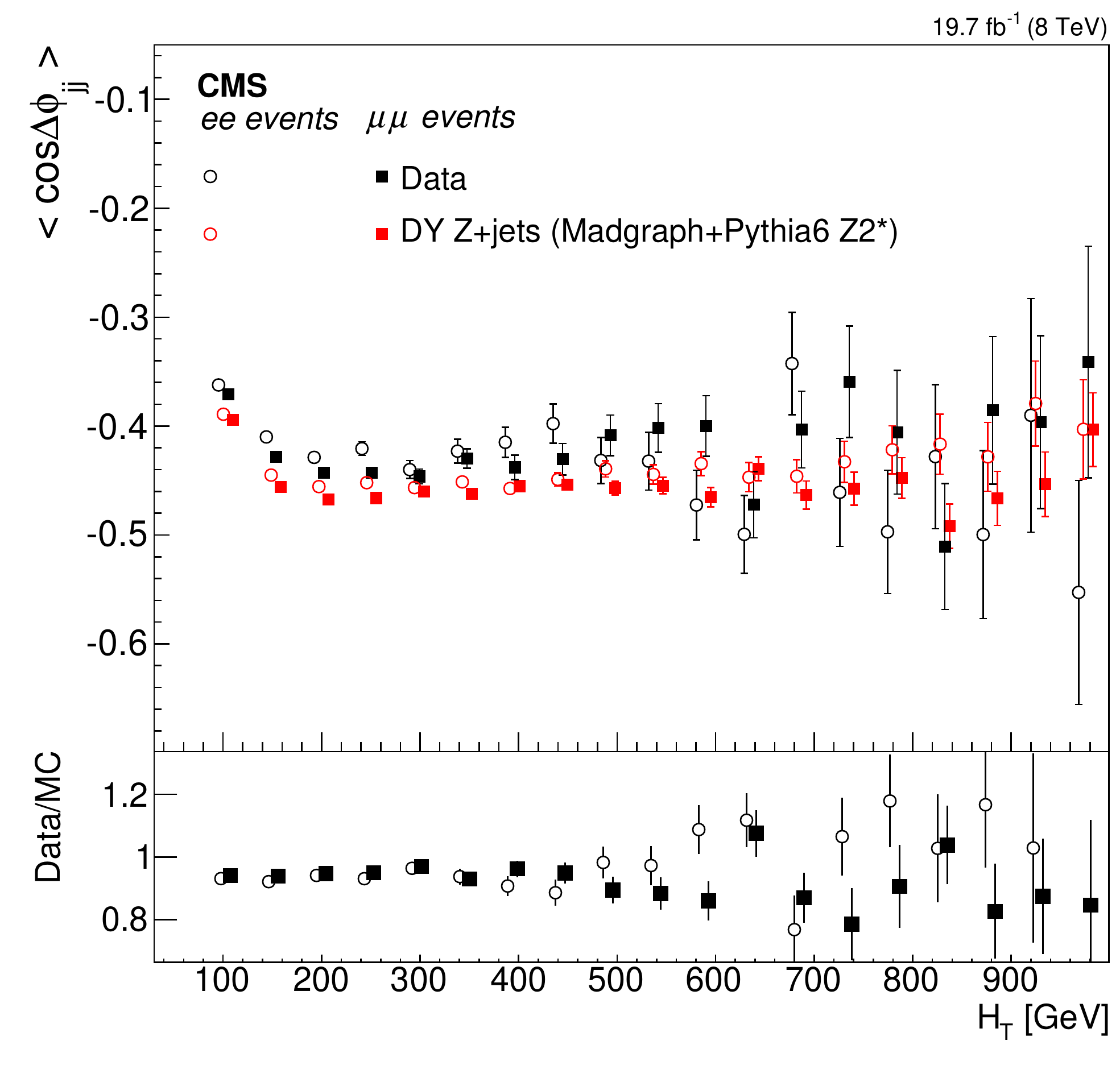}
\includegraphics[width=0.234\textwidth]{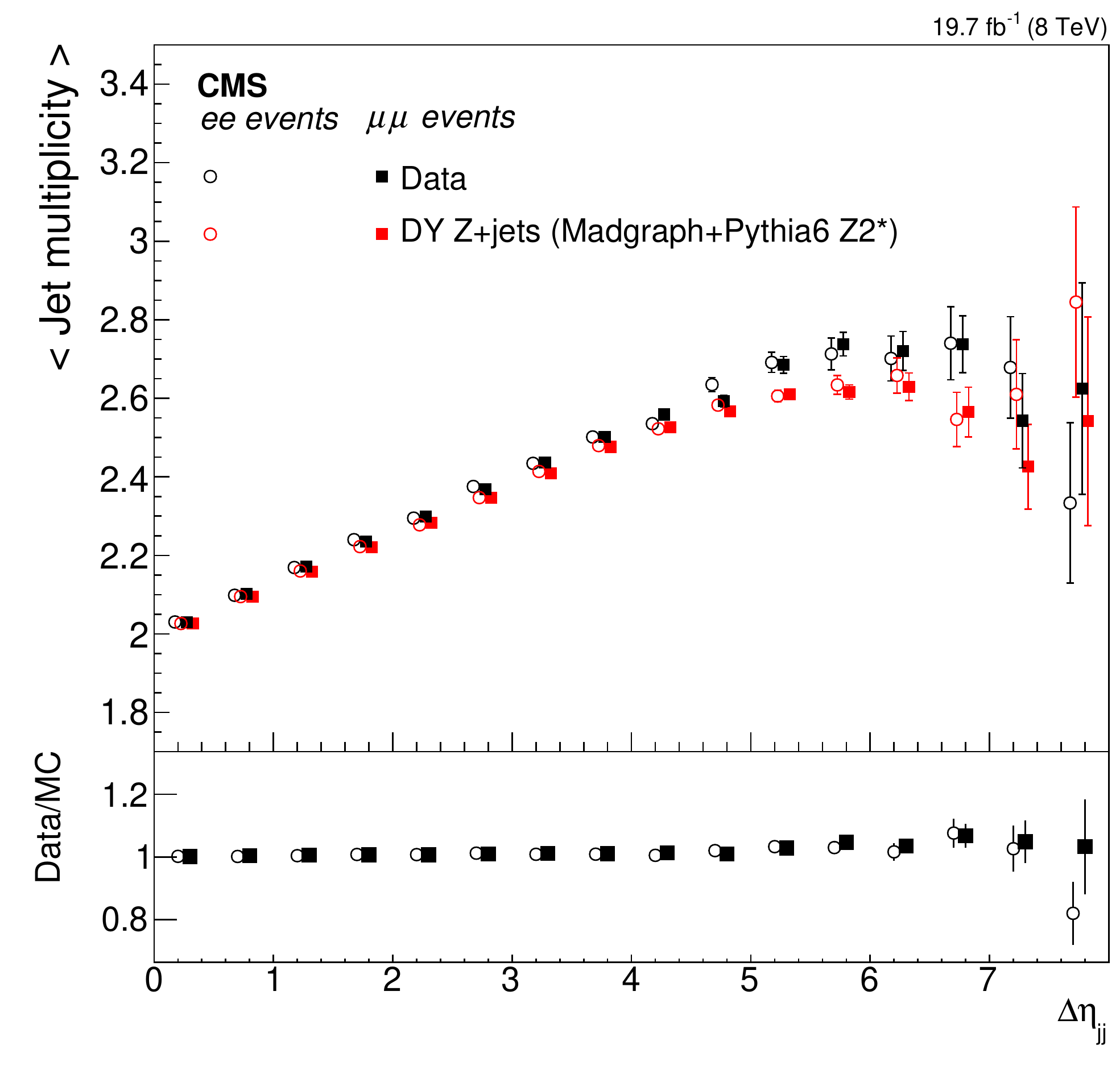}
\includegraphics[width=0.234\textwidth]{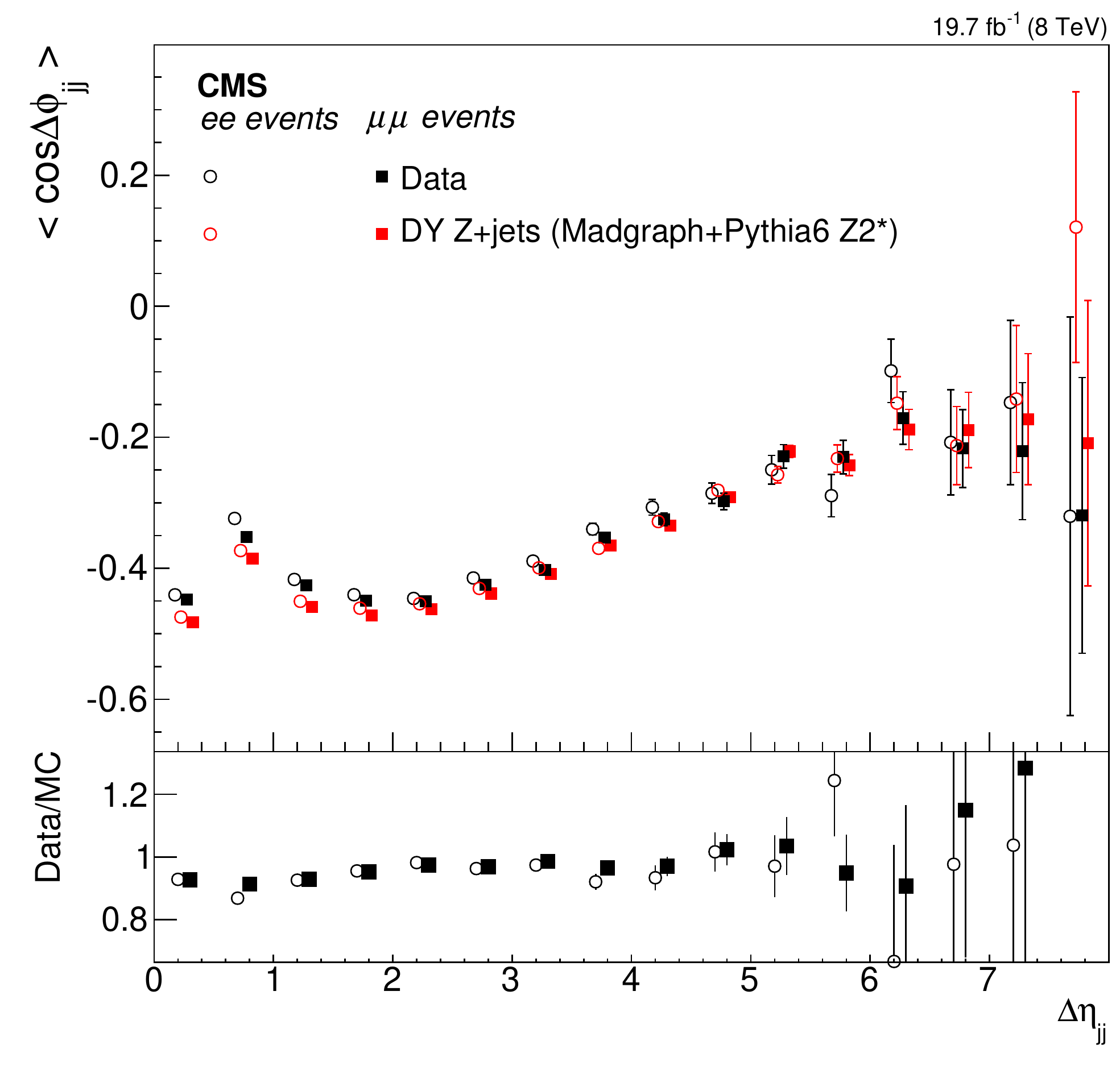}
\caption{Jet radiation patterns in selected Z plus jets events. (top left) Average number of jets with $\pt > 40$~GeV and 
(top right) average $\cos\Delta\phi_\mathrm{jj}$, as a function of the total $\HT$.
(bottom left) Average number of jets with $\pt > 40$~GeV  and 
(bottom right) average $\cos\Delta\phi_\mathrm{jj}$, as a
function of $\Delta\eta_\mathrm{jj}$ between the dijet with largest $\Delta\eta$.
The ratios of data to expectation are given below the main panels.}
\label{fig:radpat}}
\end{figure}

Figure~\ref{fig:radpat}
shows  the average number of jets and the average $\cos\Delta\phi_\mathrm{jj}$
as a function of the total $\HT$ and $\Delta\eta_\mathrm{jj}$. 
The plots indicate that 
the MADGRAPH + PYTHIA (ME+PS)
predictions are in good agreement with the data, even in the regions
of largest $\HT$ and $\Delta\eta_\mathrm{jj}$.

\section{Hadronic activity in the dijet rapidity gap}
The rapidity gap activity is studied in events with a Z and two VBF ``tagging'' PF jets 
with $\pt>50, 30$~GeV, making use of charged tracks only, and with 
additional PF jets in a region of higher signal purity. 

\begin{figure}[htb] {
\centering
\includegraphics[width=0.234\textwidth]{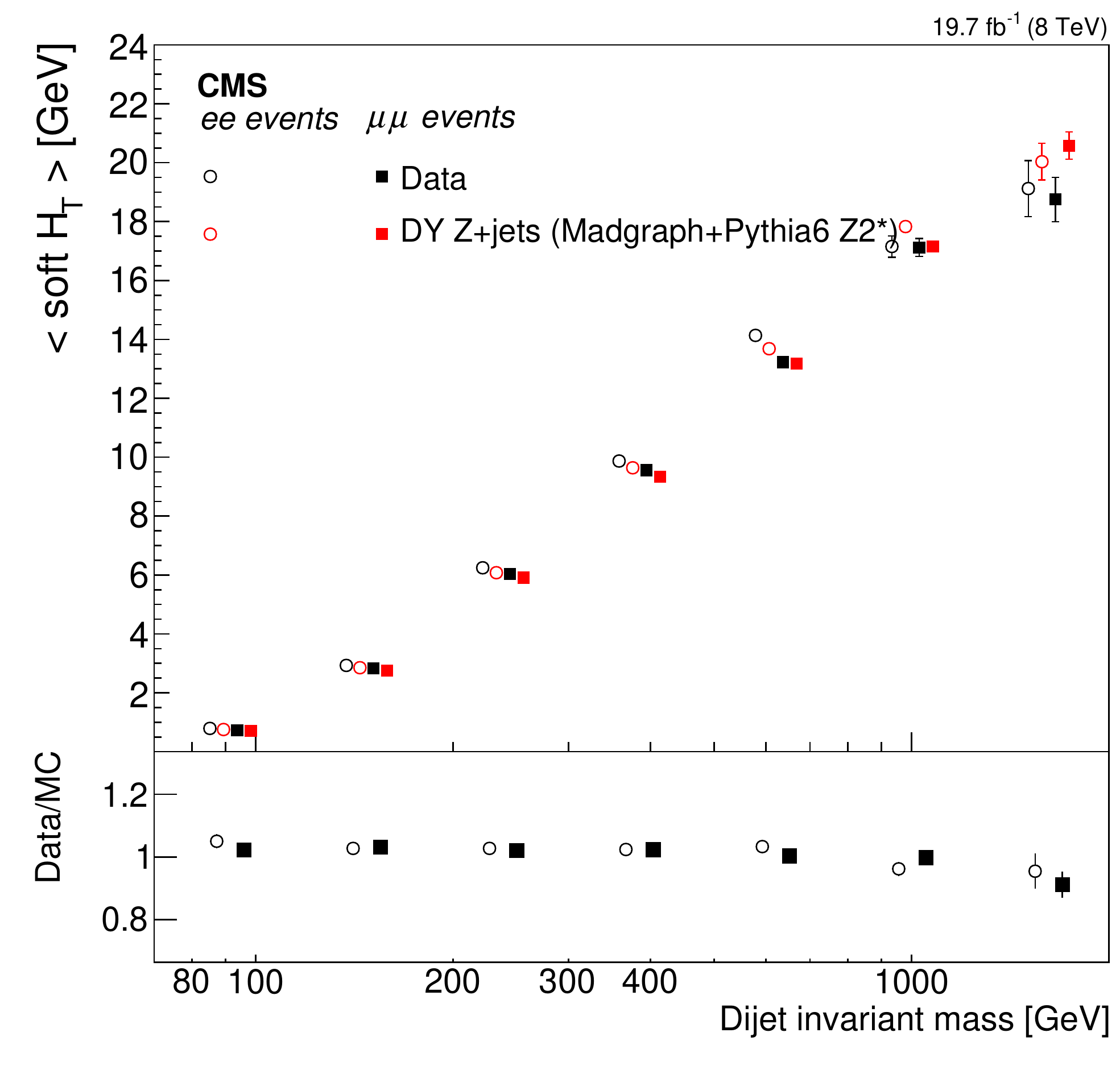}
\includegraphics[width=0.234\textwidth]{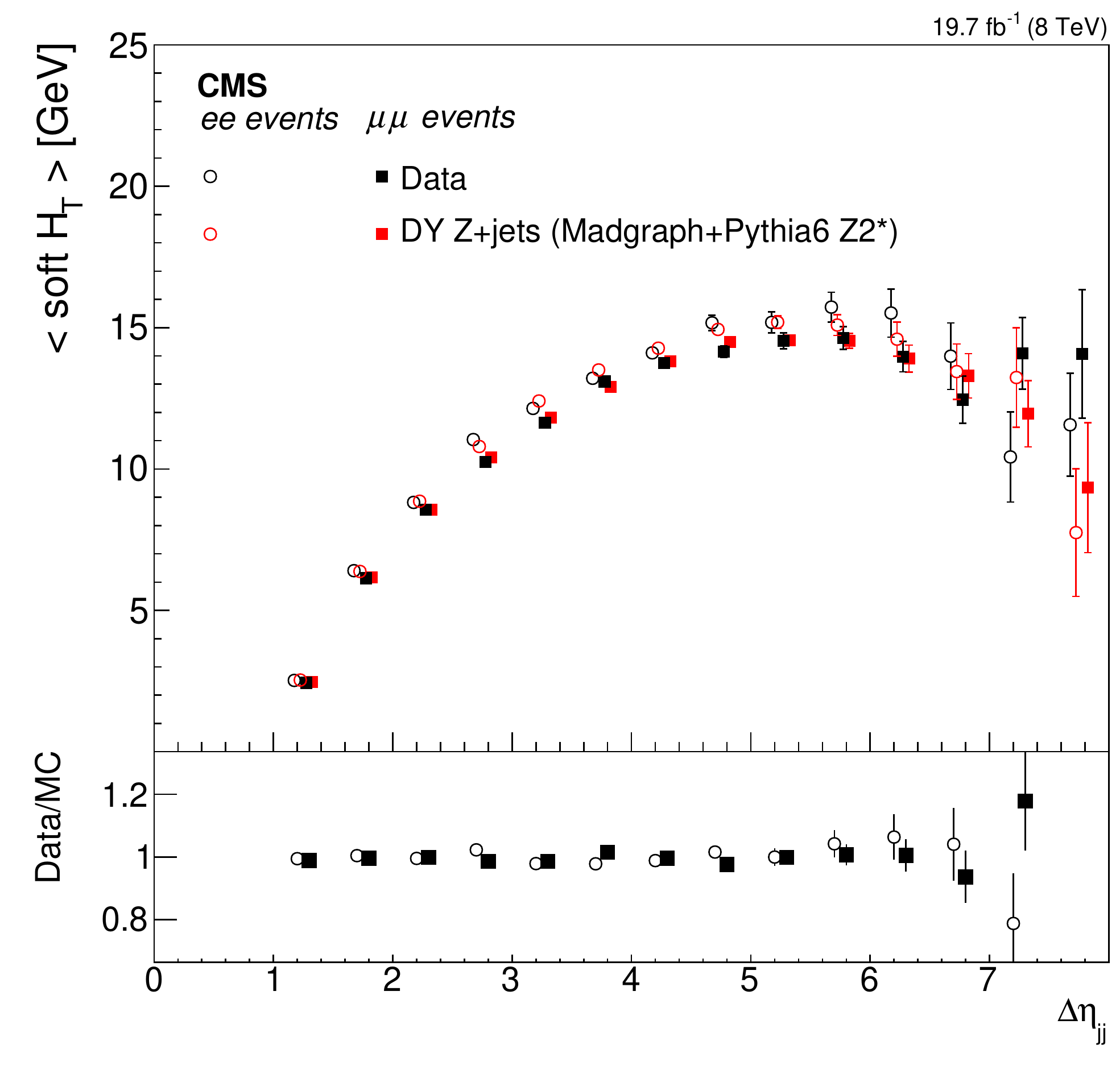}
\caption{Average soft $\HT$ computed using the three leading
soft-track jets reconstructed in the $\Delta\eta_\mathrm{jj}$
pseudorapidity interval between tagging jets with $\pt>50,30$~GeV.
The average soft $\HT$ is shown as function of:
(left) $M_\mathrm{jj}$ and (right) $\Delta\eta_\mathrm{jj}$
for both the dielectron and dimuon channels.
The ratios of data to expectation are given below the main panels.
\label{fig:softHT}}
}
\end{figure}

For the first study we use tracks associated with the main event 
primary vertex (PV), defined as the PV with
the largest $\sum \pt^2$ of the tracks used to fit the vertex, and exclude 
tracks associated with the two leptons or with the tagging jets.
A collection of ``soft track-jets'' is defined
by clustering the selected tracks using the anti-$k_{\rm T}$ algorithm
with $R=0.5$. The use of track jets represents a validated
method~\cite{CMS-PAS-JME-10-006} to reconstruct jets with energy
as low as a few GeV, that is not affected by pileup,
thanks to the PV association~\cite{CMS-PAS-JME-08-001}.
The soft $\HT$ variable is defined as the scalar sum of the \pt of up to
three leading-\pt soft-track jets in the $\eta$ interval
between the tagging jets.
The dependence of the average soft $\HT$ for selected Z plus two jet events
as a function of $M_\mathrm{jj}$ and $\Delta\eta_\mathrm{jj}$ is shown in Fig.~\ref{fig:softHT},
and good agreement is observed between data and the simulation in all ranges. 

\begin{figure}[htp]
\centering
\includegraphics[width=0.4\textwidth]{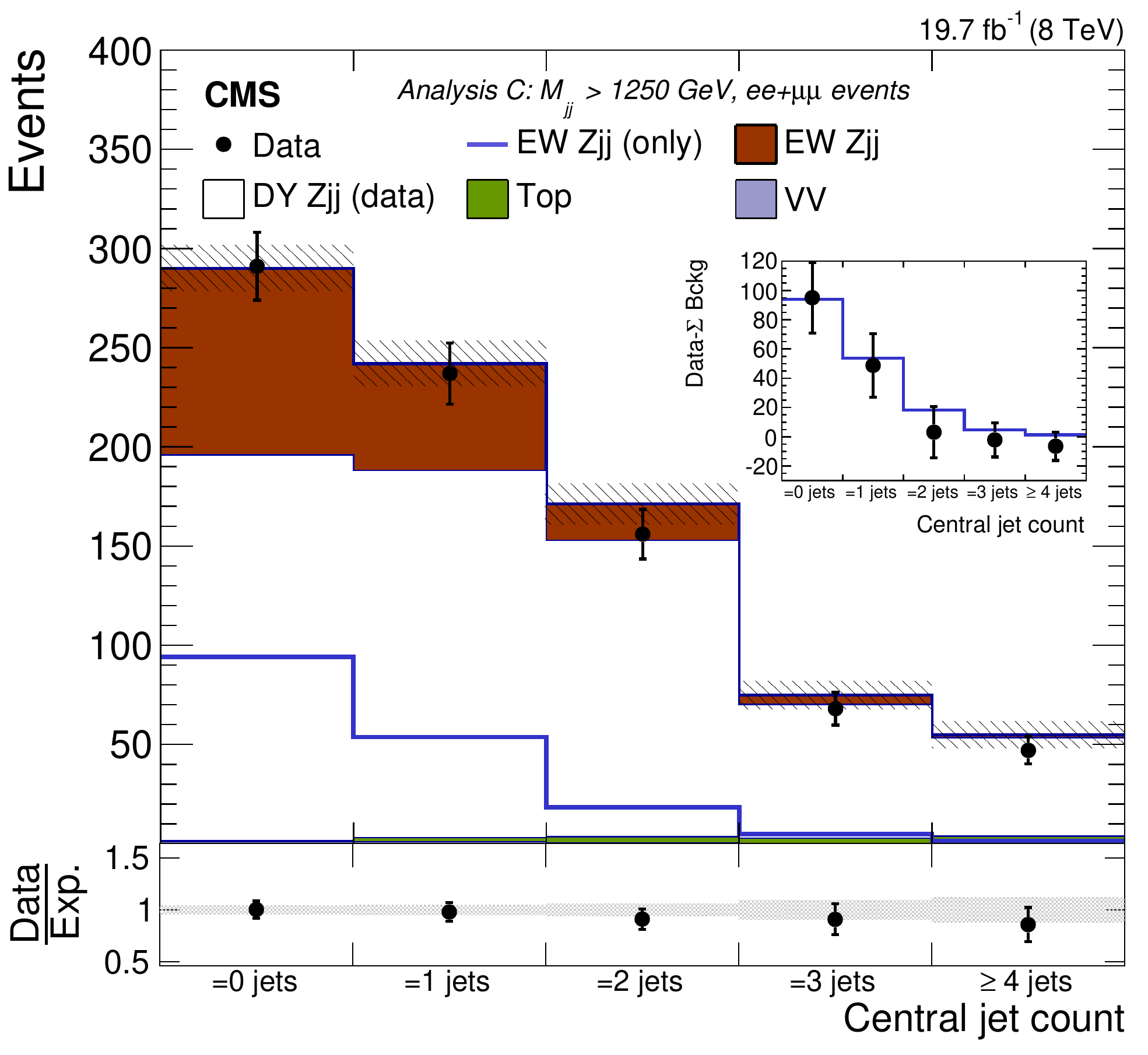}
\caption{
Additional jet multiplicity with $\pt>15$~GeV
within the $\Delta\eta_{\mathrm{jj}}$ of the two tagging jets
in events with $M_\mathrm{jj}>1250$~GeV.
In the main panels the expected contributions from signal, DY, and residual
backgrounds are shown stacked, and compared to the observed data.
The signal-only contribution is superimposed separately
and it is also compared to the residual data after the subtraction
of the expected backgrounds in the insets.
The ratio of data to expectation is represented by point markers in the bottom panels.
The total uncertainties assigned to the expectations are represented as shaded bands.}
\label{fig:rapgap}
\end{figure}

The rapidity gap interval has also been studied using PF jets with $\pt>15$~GeV,
in the $M_\mathrm{jj}>1250$~GeV region with higher signal purity, to examine 
possible evidence of the color exchange suppression for the EW signal component. 
Results are shown in Fig.~\ref{fig:rapgap} for the additional jet multiplicity 
in the dijet rapidity gap, where the data, in agreement with the MC expectations, 
indicates the presence of the EW signal with a 
suppressed third jet emission compared to the 
background-only prediction.

\begin{figure*}[htp]
\centering
\includegraphics[width=0.32\textwidth]{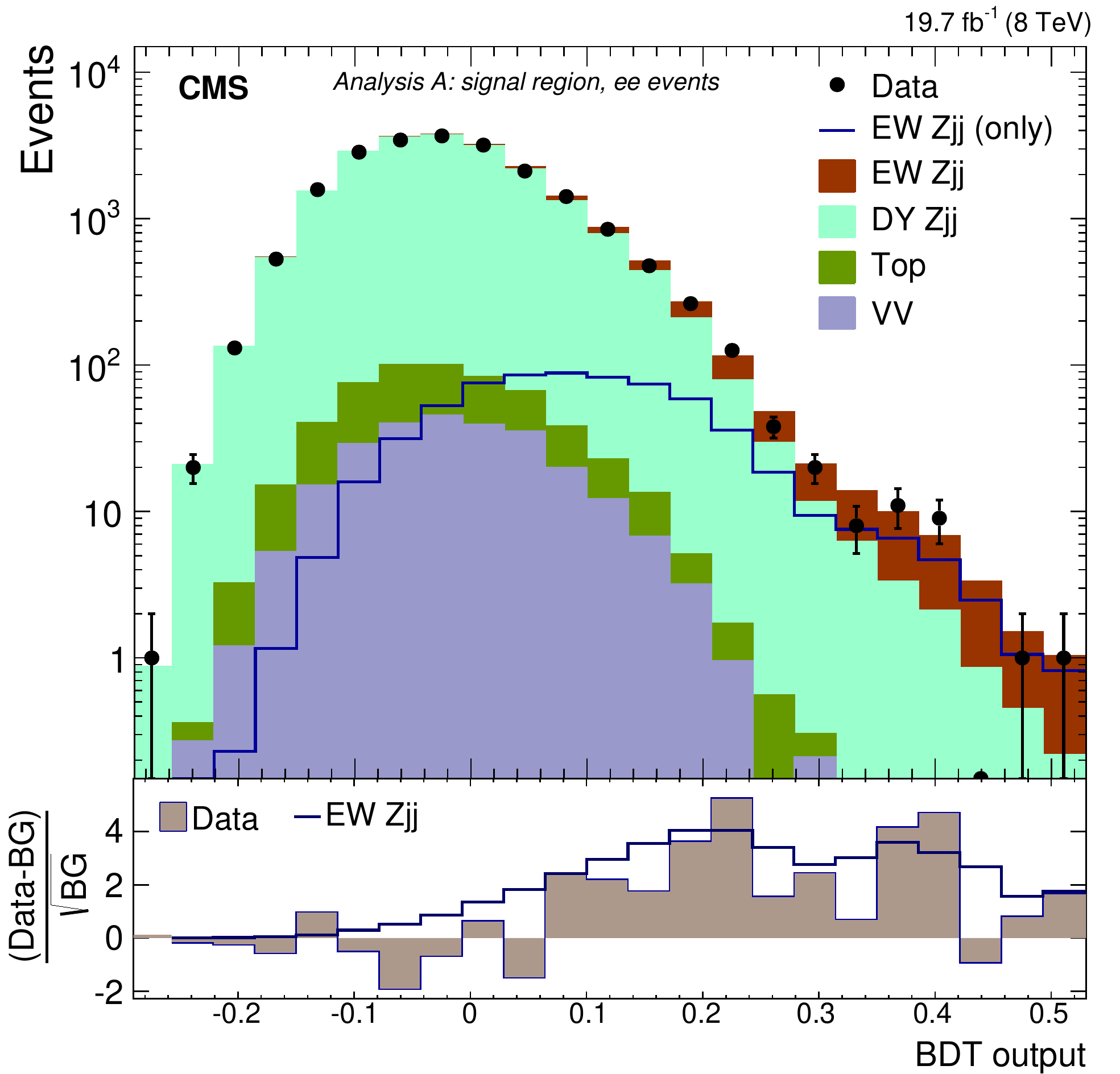}
\includegraphics[width=0.32\textwidth]{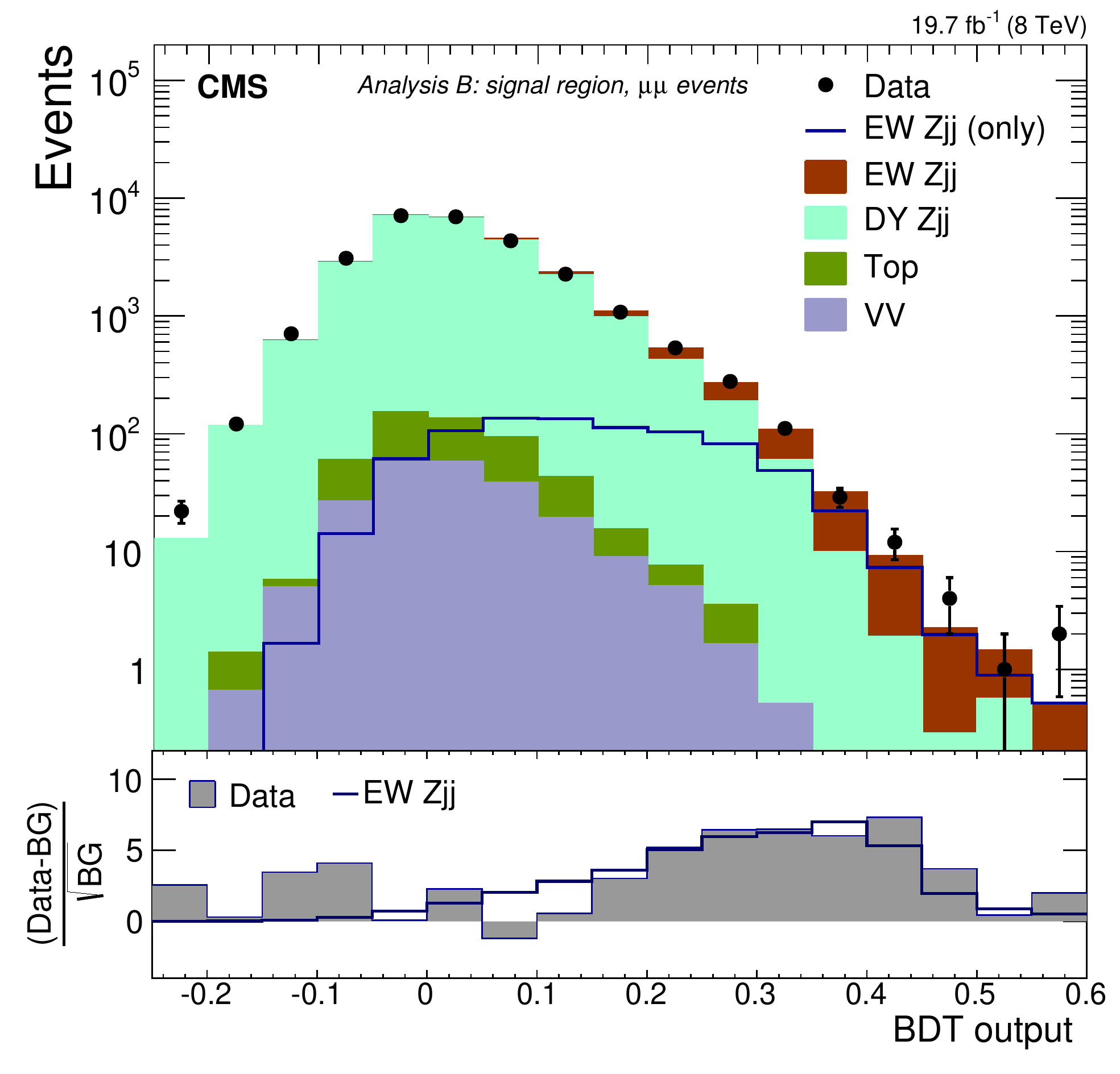}
\includegraphics[width=0.32\textwidth]{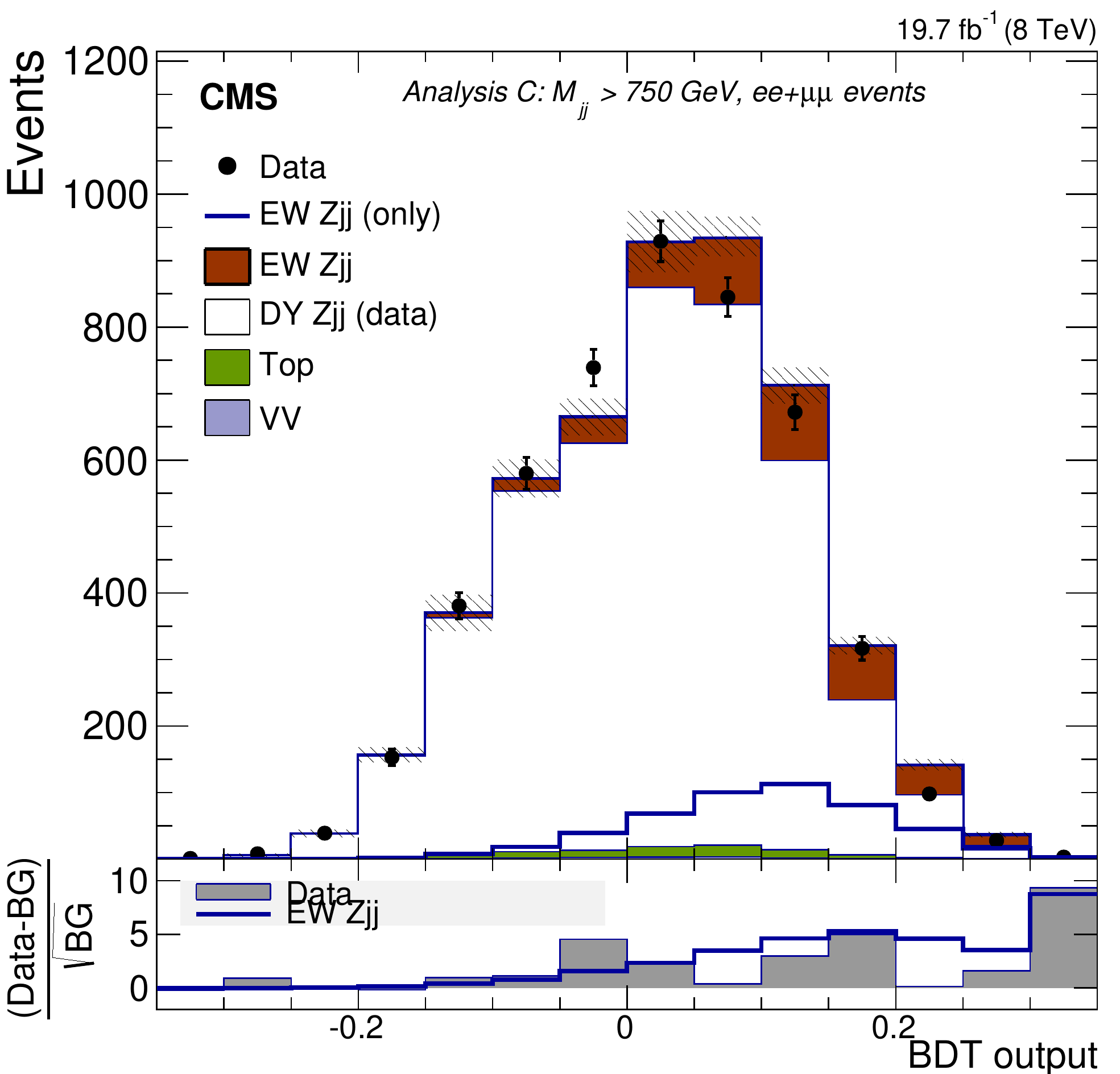}
\caption{Distributions the BDT outputs for (left) dielectron events in analysis A,
(center) dimuon events in analysis B, and all dilepton events in analysis C,
with a dijet invariant mass $M_\mathrm{jj}>750$~GeV.
The bottom panels show the differences between data or
the expected signal contribution, with respect to the expected background (BG).
}
\label{fig:bdt}
\end{figure*}

 \section{Signal measurements}
The three analyses make use of multivariate boosted decision tree (BDT) discriminators
to acquire the best expected separation of signal and background sources. 
The BDTs make mostly use of the dijet and Z boson kinematics, with the addition of 
a quark/gluon (q/g) jet discriminator~\cite{Chatrchyan:2013jya} as both signal jets are originated 
from quarks while the jets in background events are more probably 
initiated by gluons emitted from QCD processes. 

BDT output distributions for the different analyses are shown in Fig.~\ref{fig:bdt},
where a good overall agreement is observed between the data and the MC predictions.

To measure the signal cross-section, each analysis builds a binned likelihood 
based on the BDT output distributions that is used to fit strength modifiers 
for both the signal and the main DY background. 
Nuisance parameters are added to modify the
expected rates and shapes according to the estimate of the systematic
uncertainties affecting each analysis.
Possible interference effects between the signal and
the DY background processes is taken into account in the fit,
with a parameterisation derived from MADGRAPH, as a function of
the $M_\mathrm{jj}$ variable.
The statistical methodology used follows what used in CMS
Higgs analysis~\cite{Chatrchyan:2012ufa} using asymptotic formulas~\cite{Cowan:2010js}.

A summary of fitted signal strengths is reported in Table~\ref{tab:mu}, together with the 
breakdown of all relevant uncertainties.
The signal strength obtained from the combined fit of two channels in analysis A  is 
$\mu=0.84\pm 0.07\rm{(stat)} \pm0.19\rm{(syst)}$ 
corresponding to a measured signal cross section
$$\sigma({\mathrm{EW}~\ell\ell\mathrm{jj}})=\\
174\pm 15\rm{(stat)} \pm 40\rm{(syst)}\rm{fb,}$$
with the background-only hypothesis excluded with a significance
greater than 5$\sigma$.

\begin{table*}[htbp]
\centering
\caption{Measured signal strength $\mu$ in the different analyses and channels,
with statistical and relevant systematic uncertainties.\label{tab:mu}}
\begin{tabular}{l|ccc|c|ccc}\hline
& \multicolumn{3}{c|}{Analysis A} & Analysis B & \multicolumn{3}{c}{Analysis C} \\
& $\Pe\Pe$ & $\mu\mu$ & $\Pe\Pe+\mu\mu$ & $\mu\mu$ & $\Pe\Pe$  & $\mu\mu$  & $\Pe\Pe+\mu\mu$ \\
\hline
~~~Luminosity                            & 0.03 & 0.03 & 0.03 & 0.03 & 0.03 & 0.03 & 0.03\\
~~~Lepton trigger/ selection        & 0.04 & 0.04 & 0.04 & 0.04 & 0.04 & 0.04 & 0.04\\
~~~Jet Energy Scale            & 0.06 & 0.05 & 0.05 & 0.04 & 0.06 & 0.05 & 0.05\\
~~~Jet Energy Resolution                               & 0.02 & 0.02 & 0.02 & 0.02 & 0.04 & 0.04 & 0.03\\
~~~DY background                            & 0.07 & 0.05 & 0.07 & 0.08 & 0.14 & 0.12 & 0.13\\
~~~Signal acceptance                   & 0.03 & 0.04 & 0.04 & 0.04 & 0.06 & 0.06 & 0.06\\
~~~DY/EW interference           & 0.14 & 0.14 & 0.14 & 0.13 & 0.06 & 0.08 & 0.08\\\hline
Systematic uncertainty                &0.19 & 0.18 & 0.19 & 0.17 & 0.17 & 0.17 & 0.18\\\hline
Statistical uncertainty                 & 0.11 & 0.10 & 0.07 & 0.09 & 0.24 & 0.21 & 0.16 \\\hline
$\mu=\sigma/\sigma_\text{th}$   & 0.82 & 0.86 & 0.84 & 0.89 & 0.91 & 0.85 & 0.88 \\\hline
\end{tabular}
\end{table*}

\bibliographystyle{elsarticle-num}
\bibliography{azzurri}







\end{document}